\documentclass{elsart}
\usepackage{natbib, amsmath, amssymb, graphicx}

\newcommand{\mymatrix}[4]{$\left(\protect\begin{tabular}{cc} #1 & #2\protect\\ #3 & #4\protect\end{tabular}\right)$}

\begin{document}
\begin{frontmatter}

\title{Fixation of strategies for an evolutionary game in finite populations}
\author[Antal]{Tibor Antal
\thanksref{leave}},
\thanks[leave]{On leave from Institute for Theoretical Physics -- HAS, 
E\"otv\"os University, Budapest, Hungary}
\author[Scheuring]{Istv\'an Scheuring} 
\address[Antal]{Center for Polymer Studies and Department of Physics, Boston University, Boston, Massachusetts 02215, USA}
\address[Scheuring]{Department of Plant Taxonomy and Ecology, Research Group of Ecology and Theoretical Biology, E\"otv\"os University and HAS, P\'azm\'any P. s\'et\'any 1/c, H-1117, Budapest, Hungary}

\begin{abstract}
A stochastic evolutionary dynamics of two strategies given by $2\times 2$ matrix games is studied in finite populations. We focus on stochastic properties of fixation: how a strategy represented by a single individual wins over the entire population. The process is discussed in the framework of a random walk with arbitrary hopping rates. The time of fixation is found to be identical for both strategies in any particular game. The asymptotic behavior of the fixation time and fixation probabilities in the large population size limit is also discussed. We show that fixation is fast when there is at least one pure evolutionary stable strategy (ESS) in the infinite population size limit, while fixation is slow when the ESS is the coexistence of the two strategies.  
\end{abstract}
\date{\empty}

\end{frontmatter}
\section{Introduction}

Although the notion of evolutionary stable strategy (ESS) was worked out for finite populations many years ago \citep{Maynard2, Schaffer, Neill}, and there are numerous studies where finiteness and spatial structure of populations are taken into account \citep{Nowak-Sigmund}(and references therein), models of evolutionary game dynamics are generally considered in populations of infinite size \citep{Hofbauer}. 
To study evolutionary game dynamics in finite populations Nowak and coworkers \citep{Nowak, nowak04} generalized Moran's classical population genetic model \citep{Moran} by using frequency dependent fitness.    
While for infinite populations a strategy is defined to be an ESS if its fitness is greater than the fitness of any rare mutant  \citep{Maynard}, these recent works emphasize that an ESS strategy have to satisfy two requirements in finite populations: {\bf a)} it is resistant against the invasion of rare mutants and  {\bf b)} the probability that a rare mutant strategy overtakes the resident one is smaller than it is for random drift \citep{Nowak, nowak04}. In order to investigate these conditions quantitatively, Nowak determined the invasion coefficients ({\bf a)}) and the fixation probabilities ({\bf b)}) in the simplest $2 \times 2$ matrix games in the above cited works. They conclude that apart from the payoff matrix, the population size plays an equally important role in the evolutionary dynamics of finite populations. Also, by proving a set of basic theorems, these authors establish a simple classification of evolutionary scenarios in these games. Alternative definitions of ESS in finite populations are discussed in \citep{{Schaffer},{wild}}.

On the other hand, obviously, the characteristic timescales also play a crucial role in these systems.
In some situations, although fixation is probable, it might occur very slowly or rapidly \citep{Lieberman}. 
For systems with very slow fixation, the population typically consists of coexisting strategies, as it takes a long time for a strategy to win over the entire population.
Conversely, systems with short fixation times are often found to be fixated to the preferred strategy. Moreover, it is possible in principle that the two strategies, when represented by a single mutant individual, fixate with the same probability but at different speeds. In this situation the probabilities of fixation would be insufficient to describe the system. We show in this paper, however, that this latter situation never occurs.
In order to show this, we focus on the average fixation times of single mutants in the generalized Moran process introduced in \citep{nowak04}. This model is essentially a special case of a random walk with arbitrary hopping rates and two absorbing states. Some of our results are obtained for this general random walk.

The paper is organized as follows. After recalling the definition and some basic properties of the model in Sec.~\ref{model}, the exact expressions for the average time to overtake the entire population are derived for both strategies in Sec.~\ref{time}. We show that a single mutant following strategy $A$ fixates in the same average time as a single $B$ player does in a given game, although the probability of fixation for the two strategies are generally different. An alternative direct proof of this property is presented in the Appendix. In Sec.~\ref{neutral} the neutral game of identical strategies is considered as the simplest example.
In Section \ref{largen} the asymptotic behavior of the fixation time and the fixation probabilities are derived in the large population size limit. 
In Sec.~\ref{numer} we formulate three conjectures making connections between the speed of fixation and other characteristics of the game dynamics, based on numerical investigations of the derived formulas. Finally, the results are summarized and discussed in Sec.~\ref{disco}.

\section{Model and fixation probabilities}
\label{model}
  
In the evolutionary game introduced in \citep{nowak04}, the individuals can follow either $A$ or $B$ strategies defined by the payoff matrix
 \begin{equation}
\label{payoff}
\begin{tabular}{r|cc}
 & A & B \\ \hline
 A & a & b \\ 
 B & c & d \\ 
\end{tabular}
\end{equation} 
This matrix lists the average rewards of players after playing numerous times with each other. Namely 
an $A$ player receives an average reward $a$ when playing against another $A$ player, and reward $b$ when playing against a $B$ player. Similarly, a $B$ player gets reward $c$ when playing against an $A$ player and reward $d$ when playing against another $B$ player.
 
Out of the total $N$ individuals, $i$ of them follow strategy $A$, and $N-i$ follow strategy $B$. As each individual plays with every other one, the average fitness of an individual playing strategy $A$ or $B$ are proportional respectively to
\begin{equation}
\begin{split}
&f_i= a(i-1) + b(N-i)\\
&g_i= ci+d(N-i-1) ~.
\end{split}
\label{fitness}
\end{equation}
More precisely these fitness values should be normalized by $N-1$, but that can be safely ignored as only their ratio appears under the present dynamics. Then the evolution of the strategies are governed by the following Moran process \citep{Moran}, and the fitness values are recalculated after each evolutionary step. 
At each time step an individual is chosen for reproduction proportional to its fitness. This newborn individual replaces another randomly selected individual, hence the total number of the population $N$ remains constant.
Consequently, at each time step, $i$ can change at most by one according to the following transition probabilities
\begin{equation}
\begin{split}
&P(i \to i-1) = \mu_i\\
&P(i\to i) = 1-(\mu_i + \lambda_i)\\
&P(i\to i+1) = \lambda_i ~,
\label{transgen}
\end{split}
\end{equation}
where 
\begin{equation}
\begin{split}
&\mu_i = \frac{(N-i)g_i}{if_i+(N-i)g_i} \frac{i}{N}\\
&\lambda_i = \frac{if_i}{if_i+(N-i)g_i} \frac{N-i}{N}. 
\label{mu}
\end{split}
\end{equation}
Note that the ratio of the hopping probabilities is equal to the ratio of the fitnesses 
$\mu_i/\lambda_i=g_i/f_i$.
This Markov chain is essentially a random walk on sites $0\le i \le N$. 
A walk starting at site $i$ eventually hits one of the absorbing states at 0 or $N$. We denote the probability of hitting site 0 first by $\epsilon_i$ and hitting site $N$ first by $\epsilon_i^+ = 1-\epsilon_i$. 
Since in one time step the walker stays at its position or jumps to a neighboring site, these first passage probabilities obey the $N-1$ equations
\begin{equation}
\epsilon_i = \mu_i\epsilon_{i-1} + (1-\mu_i-\lambda_i)\epsilon_i + \lambda_i\epsilon_{i+1} 
\label{epsbase}
\end{equation}
for $1\le i \le N-1$, and the boundary conditions $\epsilon_0=\epsilon^+_N=1$ and 
$\epsilon_N=\epsilon^+_0=0$. The solution of these equations reads as
\begin{equation}
 \epsilon_i = \frac{s_{i,N-1}}{s_{0,N-1}}, ~~~
 \epsilon_i^+ = \frac{s_{0,i-1}}{s_{0,N-1}} ~.
 \label{epsres}
\end{equation}
where 
\begin{equation}
\label{sq}
s_{n,m}=\sum_{k=n}^m q_k ~,~~~~
q_i= \prod_{j=1}^i \frac{\mu_j}{\lambda_j} = \prod_{j=1}^i \frac{g_j}{f_j} ~,
\end{equation}
and $q_0=1$ \citep{karlin,vankampen}. 
We are particularly interested in the fixation probability for a single mutant $B$
\begin{equation}
\label{epsNres}
 \epsilon_{N-1} = \frac{q_{N-1}}{s_{0,N-1}} ~,
\end{equation}
and that of a single mutant $A$
\begin{equation}
\epsilon_1^+ = \frac{1}{s_{0,N-1}} ~.
 \label{epspres}
\end{equation}
The nature of these results are analyzed in the following sections.
A single strategist $A$ can invade $B$ if its fitness is greater than the resident $B$-s, that is if $f_1>g_1$ \citep{Maynard}. Since the probability of fixation is $1/N$ in a neutral game, selection favors fixation of a single mutant $A$ (that is $A$ to replace $B$) if $\epsilon^+_{1} > 1/N$ \citep{nowak04}. (Similar condition for the invasion of $B$ is $f_{N-1}<g_{N-1}$, and for the fixation of $B$ is $\epsilon_{N-1} > 1/N$.) Using these conditions it has been shown that there are five qualitatively different selection scenarios in finite populations for $2\times 2$ matrix games. These different scenarios are listed in Table~\ref{combi}.

\begin{table}
\begin{tabular}{lll}
Class \hspace{1cm} & Selection favors \hspace{4cm} & Speed\\
\hline
$B_{\Rightarrow \Rightarrow}^{\rightarrow\rightarrow}A$:& $A$ to invade and replace $B$ &fast or slow\\  
$B_{\Rightarrow\Leftarrow}^{\rightarrow\leftarrow}A$:& mutual invasion, $A$ and $B$ to coexist &slow\\
$B_{\Rightarrow\Rightarrow}^{\rightarrow \leftarrow}A$:& mutual invasion, $A$ to replace $B$ &fast or slow\\
$B_{\Rightarrow \Rightarrow}^{\leftarrow \rightarrow}A$:& no one to invade, $A$ to replace $B$ & fast\\
$B_{\Leftarrow \Rightarrow}^{\leftarrow \rightarrow} A$:& no one to invade or replace & fast\\
  \nonumber
\end{tabular}
\caption{The possible combinations of qualitatively different fixation and invasion events. Changing the roles of $A$ and $B$ does not alter the scenarios qualitatively.  The speed of the walk is discussed in Sec.~\ref{times}, and \ref{numer}.}
\label{combi}
\end{table}

The reason of writing $B$ in front of $A$ in the symbols of the different scenarios in Table~\ref{combi} and throughout the paper is that in this way their order is the same as that of the pure $B$ state ($i=0$) and the pure $A$ state ($i=N$) along the path of the walk. In this way the single arrows represent the preferred direction of the walk at each end of the chain. For example $B^{\rightarrow\leftarrow} A$ means that in case of a single $A$ player ($i=1$) the walk prefers to step to the right (first arrow), but with a single $B$ player ($i=N-1$) it prefers to step to the left (second arrow). The double arrows point to the direction in which the absorption is more possible than in the neutral game. As an example $B_{\Rightarrow \Rightarrow}A$ means that a system with a single $A$ player (first arrow) and also a system with a single $B$ player (second arrow) ends up in a pure $A$ state more probably than in a neutral game. 

\section{Average fixation time}
\label{time}

A walk starting at site $i$ eventually gets absorbed into either site 0 or site $N$. We are interested in the mean conditional exit (or absorption) time $t_i$ to site 0, and $t_i^+$ to site $N$. The problem was formulated and solved for the simple random walk in \citep{{landauer},{fisher}} in a more general framework. The solution for a random walk with arbitrary hopping probabilities is given in \citep{vankampen}. For completeness we present here an elementary derivation using standard methods.

Let $P_i(t)$ be the probability that the walk gets absorbed into site 0 at time $t$, if starts at site $i$ at time 0. Obviously $\epsilon_i = \sum_{t=0}^\infty P_i(t)$,
and we use the convention $P_0(t)=\delta_{t,0}$, and $P_N(t)\equiv 0$. By definition, the conditional average exit time is
\begin{equation}
 t_i = \frac{\sum\limits_{t=0}^\infty t P_i(t)}{\sum\limits_{t=0}^\infty P_i(t)} 
 = \frac{1}{\epsilon_i} \sum_{t=0}^\infty t P_i(t) ~.
\end{equation}
For $P_i(t)$ the following recursive equation holds
\begin{equation}
 P_i(t) = \mu_i P_{i-1}(t-1) + (1-\mu_i - \lambda_i)P_i(t-1) + \lambda_i P_{i+1}(t-1)
\end{equation}
for $1\le i \le N-1$. As we are interested in the first moment of $P_i(t)$, we multiply both sides of this equation by $t$, and then sum it up from $t=0$ to $\infty$. We also apply the identity
\begin{equation}
\sum_{t=0}^\infty t P_i(t-1) = \sum_{t=0}^\infty (t+1) P_i(t) = \epsilon_i(t_i+1) ~,
\end{equation}
[where $P_i(t)=0$ for $t<0$] to arrive at
\begin{equation}
\epsilon_i t_i = \mu_i \epsilon_{i-1}(t_{i-1}+1) + (1-\mu_i - \lambda_i)  \epsilon_{i}(t_{i}+1)
+ \lambda_i \epsilon_{i+1}(t_{i+1}+1) ~.
\end{equation}
Using Eq.~(\ref{epsbase}), and the notation $\tau_i=\epsilon_i t_i$, the above equation simplifies to 
\begin{equation}
-\epsilon_i = \mu_i \tau_{i-1} - (\mu_i+\lambda_i) \tau_{i}  + \lambda_i \tau_{i+1} ~,
\label{ttaurec}
\end{equation}
which, in terms of $\sigma_i=\tau_i-\tau_{i+1}$, becomes
\begin{equation}
 \sigma_i = \frac{\mu_i}{\lambda_i} \sigma_{i-1} + \frac{\epsilon_i}{\lambda_i} ~.
\end{equation}
Iterating this relation gives
\begin{equation}
 \sigma_i = \sigma_0 q_i + q_i \sum_{k=1}^i \frac{\epsilon_k}{\lambda_k q_k} ~.
 \label{tagree}
\end{equation}
Since $\sigma_0=-\tau_1$, we know all $\tau_i$ for $1\le i\le N-1$ as a function of $\tau_1$
\begin{equation}
 \tau_i = \tau_1 - \sum_{n=1}^i \sigma_n ~.
 \label{taui}
\end{equation}
The value of $\tau_1$ can be obtained from equation (\ref{ttaurec}) with $i=N-1$,
noting that $\tau_N=0$,
\begin{equation}
 \tau_1 = (1-\epsilon_1) \sum\limits_{n=1}^{N-1}q_n  \sum_{k=1}^n \frac{\epsilon_k}{\lambda_k q_k} ~,
\end{equation}
hence the conditional average time ($t_i=\tau_i/\epsilon_i$) to reach the origin is given by
\begin{equation}
 \tau_i = (1-\epsilon_i) Q_N  - Q_i ~,
 \label{taufinal}
\end{equation}
where we used the shorthand notation
\begin{equation}
Q_i = \sum_{n=1}^{i-1} q_n \sum_{k=1}^n \frac{\epsilon_k}{\lambda_k q_k} ~.
\label{Qdef}
\end{equation}
The conditional average time $t_i^+$ to reach the other absorbing state at site $N$ can be obtained similarly through $\tau_i^+=\epsilon_i^+ t_i^+$
\begin{equation}
 \tau_i^+ = (1-\epsilon_i^+) R_0  - R_i ~,
 \label{taupfinal}
\end{equation}
with
\begin{equation}
R_i = \sum_{n=i+1}^{N-1} q_{n-1} \sum_{k=n}^{N-1} \frac{\epsilon_k^+}{\lambda_k q_{k}} ~.
\end{equation}
Formula (\ref{taupfinal}) can be also obtained directly from (\ref{taufinal}) by noting that to arrive from site $i$ to the origin in a given chain is the same as to arrive from site $N-i$ to site $N$ in a mirrored chain, where all $\lambda_j$ are changed to $\mu_{N-j}$ and all $\mu_j$ are changed to $\lambda_{N-j}$. 

The expressions for the average exit times are greatly simplified for the biologically relevant $\tau_{N-1}$ and $\tau_1^+$. We can rewrite formula (\ref{taufinal}) as
\begin{equation}
  \tau_{N-1} = (1-\epsilon_{N-1})Q_N - Q_{N-1} = q_{N-1}
  \sum_{k=1}^{N-1}\frac{\epsilon_k}{\lambda_k q_k} - \epsilon_{N-1}Q_N ~,
\end{equation}
and substituting expression (\ref{epsres}) into $\epsilon_k$ we arrive at our final formula
\begin{equation}
 t_{N-1} = \sum_{n=1}^{N-1} \frac{s_{0,n-1}s_{n,N-1}}{\lambda_n q_n s_{0,N-1}}  ~.
 \label{tfixfinal}
\end{equation}
Following the same steps for (\ref{taupfinal}) leads to an expression for $t_1^+$ 
which is identical to (\ref{tfixfinal}), thus we call it the fixation time
\begin{equation}
 t_\mathrm{fix} (N)
\equiv  t_{N-1} = t_1^+ ~,
\label{equal}
\end{equation}
This surprising result states that the mean conditional exit times starting from one side and being absorbed into the other side of the chain does not depend on the particular side the walk starts, for {\it arbitrary} positive transition probabilities (it is obviously necessary that all $\mu_i>0$ and $\lambda_i>0$). 
We also support this intriguing property of the walk by a simple argument in the Appendix. 
In the special case of random walks with {\it constant bias} ($\mu_i=1-\lambda_i=$~const), the exit times are symmetric for all initial positions $t_{N-i} = t_i^+ $ as shown in \citep{fpp}.
Note also that for the simple random walk ($\mu_i=\lambda_i=1/2$) the formulas (\ref{taufinal}) and (\ref{taupfinal}) simplify to \citep{{fisher},{fpp}}
\begin{equation}
\label{simpletime}
t_i=\frac{i}{3}(2N-i)~~,~~~ t_i^+=\frac{1}{3}(N^2-i^2) ~.
\end{equation}

\subsection{Fixation time in the neutral game}
\label{neutral}

In the context of the game theory, the transition probabilities $\mu_i$ and $\lambda_i$ depend on the fitness of individuals using a given strategy, as it is given by Eq.~(\ref{transgen}). A game is given by the payoff matrix (1), 
and the fitness of individuals using strategy A and strategy B are determined by Eqs. (\ref{fitness}).  
The simplest game is the neutral game given by the payoff matrix
\begin{equation}
\label{payoff2}
\begin{tabular}{r|cc}
 & A & B \\ \hline
 A & 1 & 1 \\ 
 B & 1 & 1 \\ 
\end{tabular}
\end{equation} 
and the corresponding transition probabilities are
\begin{equation}
  \mu_i = \lambda_i = \frac{i(N-i)}{N^2} .
\end{equation}
Hence the exit probabilities in a neutral game coincide with that of the simple (fair) random walk
\begin{equation}
  \epsilon^+_1 =  \epsilon_{N-1} = \frac{1}{N} ~.
\end{equation}
On the other hand, as $\mu_i+\lambda_i<1$, the conditional average exit times are different. Using Eq.(\ref{tfixfinal}), after some straightforward algebra we arrive at
\begin{equation}
  t_\mathrm{fix}(N) = t_{N-1} = t_1^+ = N(N-1) ~,
\end{equation}
that is a neutral game fixates three times slower than a simple random walk (\ref{simpletime}) for large $N$.

It seems natural to characterize the average "speed" of fixation in a non-neutral game
by comparing the actual fixation time $t_\mathrm{fix}(N)$ to that in a neutral game $N(N-1)$. 
If $t_\mathrm{fix}(N) >N(N-1)$ then fixation is said to be \emph{slow}, if $t_\mathrm{fix}(N) < N(N-1)$ then fixation is \emph{fast}. It is easy to check that for an $N=2$ system $t_\mathrm{fix}(2)=2$ independently of the elements of the payoff matrix. To write the fixation time and probability in a closed form for a general payoff matrix and for arbitrary $N$ seems to be a hopeless task, but they can be derived asymptotically in the large $N$ limit.

\section{Large $N$ limit}
\label{largen}

In this section we derive the large $N$ asymptotic behavior of the exact expressions for the fixation probabilities Eq.(\ref{epsNres}-\ref{epspres}) and times Eq.(\ref{tfixfinal}). What large means depends on the actual payoff matrix; the speed of convergence can be slow for some particular choice of matrix elements (See Fig.~\ref{coex} and \ref{domin}). We emphasize that the large $N$ behavior of a finite chain is different from the behavior in an infinite population \citep{Neill}, modeled either by a walk on a semi-infinite chain or by a replicator equation description \citep{nowak04}.

Let us derive the asymptotic behavior of the basic quantity $q_k$ first 
\begin{equation}
  q_k = \prod_{i=1}^k \frac{g_i}{f_i} = \exp\left(\sum_{i=1}^k \ln \frac{g_i}{f_i} \right) ~.
\end{equation}
When $N$ is large, the above sum can be written as an integral
\begin{equation}
  N\int_0^y dx~ \ln \frac{x(c-d)+d}{x(a-b)+b} ~,
\end{equation}
where $y=k/N$. Evaluating the integral we arrive at  $q_k = \tilde q(y)^N$, with
\begin{equation}
\label{qtilde}
  \tilde q(y) = \left( \frac{d}{b}\right)^y \frac{\left(1+y\frac{c-d}{d}\right)^{y+\frac{d}{c-d}}}
  {\left(1+y\frac{a-b}{b}\right)^{y+\frac{b}{a-b}}} ~.
\end{equation}
The function $\tilde q(y)$ is remarkably simple: it is either convex if $a-c<b-d$, or concave if $a-c>b-d$.
At the boundaries
\begin{equation}
 \tilde q(0) = 1 ~,~~ \tilde q(1) = 
 \frac{d\left( \frac{c}{d}\right)^\frac{c}{c-d}}{b\left( \frac{a}{b}\right)^\frac{a}{a-b}} ~.
\end{equation}
Under the interchange of players ($A\leftrightarrows B$, that is $a\leftrightarrows d$ and $b\leftrightarrows c$) the function has the symmetry
\begin{equation}
\label{qchplayers}
 \tilde q_{A\leftrightarrows B}(y) = \frac{\tilde q(1-y)}{\tilde q(1)} ~.
\end{equation}

Since $q_k = \tilde q(y)^N$, the main contribution to the basic quantity $s_{0,N-1}$ comes from the $q_k$ terms with $k/N\approx y_\mathrm{max}$, where $\tilde q(y)$ takes its maximum at $y_\mathrm{max}$. 
According to the value of $y_\mathrm{max}$ there are four scenarios, in agreement with the replicator equation description \citep{nowak04}
\begin{enumerate} 
\item $B\longrightarrow A$: $A$ dominates $B$, if $a>c$ and $b>d$; $y_\mathrm{max}=1$
\item $B\longleftarrow A$: $B$ dominates $A$, if $a<c$ and $b<d$; $y_\mathrm{max}=0$
\item $B\leftarrow\rightarrow A$: $A$ and $B$ are bi-stable, if $a>c$ and $b<d$; $y_\mathrm{max}=y^*$ 
\item $B\rightarrow\leftarrow A$: $A$ and $B$ coexist, if $a<c$ and $b>d$; $\tilde q(y)$ has minimum at $y^*$, and its maximum is at $y_\mathrm{max}=0$ for $1>\tilde q(1)$, while its maximum is at  $y_\mathrm{max}=1$ for $1<\tilde q(1)$, 
\end{enumerate}
where  
\begin{equation}
\label{ystar}
 y^* = \frac{d-b}{a-c+d-b} ~.
\end{equation}
This suggests that the relevant parameters of the large $N$ behavior are $a-c$ and $b-d$, hence we shell depict our results as a function of these variables in Fig.~\ref{diagprob} and \ref{diagtime}.
Again the arrows in symbols represent the preferred direction of the walk at each ends of the chain.
$B\longrightarrow A$ can be thought of as a short form of $B\rightarrow\rightarrow A$.

Using the $N\to\infty$ limit in expression (\ref{qtilde}) we find that around the ends of the chain
\begin{equation}
\label{qkapprox}
q_k \approx \left\{
\begin{tabular}{ll}
$\left(\frac{d}{b} \right)^k$ & , for  $k\ll N$ ~,\\
     $\tilde q(1)^N \left(\frac{a}{c} \right)^{N-k}$ & , for $N-k\ll N$ ~.
\end{tabular} \right.
\end{equation}

\subsection{Fixation probabilities}

In order to calculate the fixation probabilities in the $N\to\infty$ limit, we need to know the asymptotic form of $s_{0,N-1}$. Since $q_k = \tilde q_k^N$, we only need the terms around the maximum of $\tilde q(y)$. In the case $B\longrightarrow A$ (that is $a>c$ and $b>d$), the maximum is at $y=0$, hence, using Eq.~(\ref{qkapprox})
\begin{equation}
 s_{0,N-1}=\sum_0^{N-1}q_k \approx \sum_0^\infty \left(\frac{d}{b}\right)^k = \frac{b}{b-d} ~.
\end{equation}
This makes the fixation probability in the $N\to\infty$ limit
\begin{equation}
\label{eps1+first}
 \epsilon_1^+ = \frac{1}{s_{0, N-1}} = 1-\frac{d}{b} ~.
\end{equation}
Surprisingly, as this result is valid as long as the maximum of $\tilde q(y)$ is at $y_\mathrm{max}=0$, this is the fixation probability in the case of  $B\rightarrow\leftarrow A$ (that is $a<c$ and $b>d$) when additionally $\tilde q(1)<1$. This means that even when $A$ and $B$ coexist, the fixation probability can be finite, even in the $N\to\infty$ limit. In this latter case however, as we shall see, the fixation time is exponentially large. In the above two cases the fixation probability for the other player becomes exponentially small
\begin{equation}
\label{epsN-1first}
 \epsilon_{N-1} = \frac{q_{N-1}}{s_{0, N-1}}  \sim \tilde q(1)^N ~,
\end{equation}
where we used Eq.~(\ref{qkapprox}). While expression (\ref{eps1+first}) for $\epsilon_1^+$ becomes exact as $N\to\infty$, in the expression (\ref{epsN-1first}) for $\epsilon_{N-1}$, we do not know the coefficient of $q(1)^N$ exactly. In other words we only know that $\ln(\epsilon_{N-1})/N\to\ln\tilde q(1)$ as $N\to\infty$.
This is comprehensible as we derived the leading order behavior in both cases.

Similarly, for $B\longleftarrow A$ (that is $a<c$ and $b<d$), and also for $B\rightarrow\leftarrow A$ when $\tilde q(1)>1$, the maximum of $\tilde q(y)$ is at $y_\mathrm{max}=1$, and repeating the same steps as before we arrive at
\begin{equation}
\epsilon_1^+ =  \left(1-\frac{a}{c}\right) \tilde q(1)^{1-N} \sim \tilde q(1)^{-N}
~,~~~ \epsilon_{N-1} = 1-\frac{a}{c}
\end{equation}

In the remaining region, $B\leftarrow\rightarrow A$ (that is $a>c$ and $b<d$), the main contribution to $s_{0,N-1}$ comes from the terms around $y_\mathrm{max}=y^*$, where $\tilde q(y)$ takes its maximum $q^*=\tilde q(y^*)>1$. In the $N\to\infty$ limit the sum $s_{0,N-1}$ can be replaced by an integral, which can be approximated by a Gaussian integral (steepest descent method)
\begin{equation}
  s_{0,N-1} \approx Nq^{*N}\int_{-\infty}^\infty dy \exp\left( -\frac{(y-y^*)^2}{2\sigma^2} \right)
  \sim q^{*N} \sqrt N ~.
\end{equation}
Thus, in this case both probabilities are exponentially small in the $N\to\infty$ limit
\begin{equation}
 \epsilon_1^+ \sim \frac{1}{\sqrt N q^{*N}} ~,~~~
 \epsilon_{N-1} \sim \frac{1}{\sqrt N} \left( \frac{\tilde q(1)}{q^*}\right)^N ~.
\end{equation}

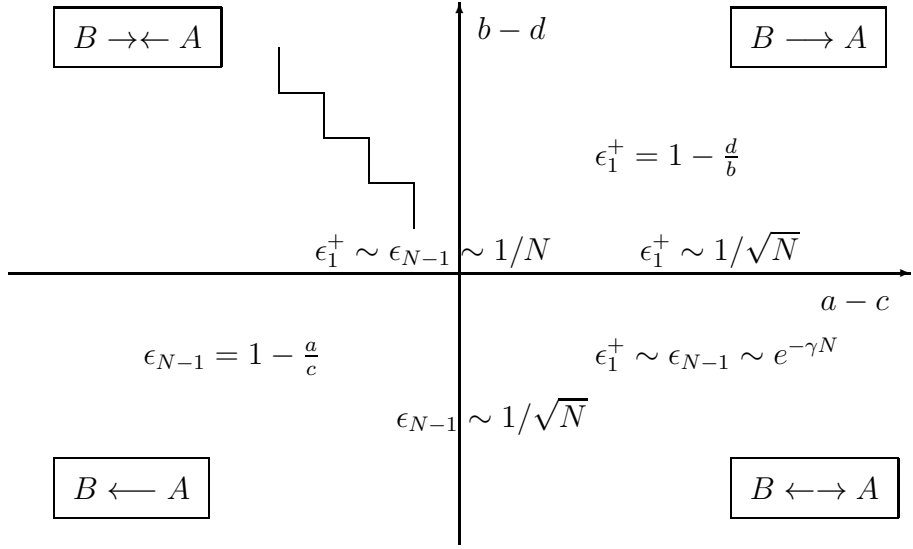
\begin{figure}
\setlength{\unitlength}{1.2mm}
\begin{center}
\begin{picture}(100,60)(0,0)
\put(0,30){\vector(1,0){100}}
\put(90,26){$a-c$}
\put(50,0){\vector(0,1){60}}
\put(52,56){$b-d$}
\put(30,55){\line(0,-1){5}}
\put(30,50){\line(1,0){5}}
\put(35,50){\line(0,-1){5}}
\put(35,45){\line(1,0){5}}
\put(40,45){\line(0,-1){5}}
\put(40,40){\line(1,0){5}}
\put(45,40){\line(0,-1){5}}
\put(5,55){\framebox{$B\rightarrow\leftarrow A$}}
\put(5,5){\framebox{$B\longleftarrow A$}}
\put(80,55){\framebox{$B\longrightarrow A$}}
\put(80,5){\framebox{$B\leftarrow\rightarrow A$}}
\put(15,20){$\epsilon_{N-1} = 1-\frac{a}{c}$}
\put(65,42){$\epsilon^+_1 = 1-\frac{d}{b}$}
\put(65,20){$\epsilon^+_1\sim\epsilon_{N-1}\sim e^{-\gamma N}$}
\put(34,31.5){$\epsilon^+_1\sim\epsilon_{N-1}\sim 1/N$}
\put(43,13){$\epsilon_{N-1}\sim 1/\sqrt N$}
\put(70,31.5){$\epsilon^+_1\sim 1/\sqrt N$}
\end{picture}
\end{center}
\caption{Graphical illustration of the different scenarios for the fixation probabilities in the large $N$ limit as summarized in Table~\ref{sum} (the value of $\gamma>0$ can be read out from Table~\ref{sum}). $\epsilon^+_1$ and $\epsilon_{N-1}$ are exponentially small where not specified. The zig-zag line in the $B\rightarrow\leftarrow A$ region refers to the boundary given by $\tilde q(1)=1$, which is not a line in these variables. The probabilities in this region are equal to those in one of the neighboring regions.}
\label{diagprob}
\end{figure}

To complete our investigation of the large $N$ behavior of the absorption probabilities we focus now on the axes (see Fig.~\ref{diagprob}), that is when either $a=c$ or $b=d$. When both are equal, then $g_i/f_i\approx q_i\approx 1$, and we asymptotically recover the behavior of the neutral game, $\epsilon^+_1\sim\epsilon_{N-1}\sim 1/N$. When $a=c$ and $b>d$, then the maximum of $\tilde q(y)$ is at 0 with non-zero derivative, thus the probabilities are the same as in the neighboring $B\longrightarrow A$ region and in the $B\rightarrow\leftarrow A$ region for $\tilde q(1)<1$. The same scenario occurs for $b=d$ and $a<c$ as it can be obtained by interchanging the players.

Significantly different behavior is found on the other halves of the axes. For $a=c$ and $b<d$ the maximum of $\tilde q(y)$ is at 1 but with zero derivative. Hence we should use again the steepest descent method to obtain $s_{0,N-1}\sim\tilde q(1)^N \sqrt{N}$ which results in $\epsilon^+_1 \sim q(1)^{-N}/\sqrt{N}$ and $\epsilon_{N-1}\sim1/\sqrt{N}$. The slowly decaying $\epsilon^+_1$ on this axes interplays between the fast decaying behavior in the $B\leftarrow\rightarrow A$ region and the constant probability in the $B\longrightarrow A$ region. By interchanging the players it immediately follows that for $b=d$ and $a>c$, $\epsilon^+_1 \sim 1/\sqrt{N}$ and $\epsilon_{N-1}\sim q(1)^{N}/\sqrt{N}$.

These results are summarized in Table~\ref{sum} and depicted in Fig.~\ref{diagprob}.

\subsection{Fixation times}
\label{times}

To derive the large $N$ asymptotic we use the exact expression for the fixation time $t_\mathrm{fix}$ given by Eq.~(\ref{tfixfinal}). Consider first the $B\longrightarrow A$ case. Here, the main contribution to $t_\mathrm{fix}$ comes from the terms around $y=0$ and $y=1$, where $1/\lambda_i$ develops singularities. Indeed in the $N\to\infty$ limit, $\lambda_i$ of Eq.(\ref{mu}) can be written as
\begin{equation}
 \lambda_i = \frac{y(1-y)}{y+(1-y)\beta(y)} ~,
\end{equation}
where $\beta(y=i/N)=g_i/f_i$, and it behaves asymptotically as
\begin{equation}
 \lambda_i = \left\{
 \begin{split}
 \frac{d}{b} ~\frac{1}{y}  &  \quad \mbox{ for }  y\ll 1\\
 \frac{1}{1-y} &  \quad \mbox{ for } 1-y\ll 1 ~.\\ 
 \end{split}
 \right.
\end{equation}
Let us also investigate the asymptotic behavior of the other terms in Eq.~(\ref{tfixfinal}).
As the main contribution to $s_{0,N-1}$ comes from the first few terms, the ratio $s_{0,n-1}/s_{0,N-1}\to1$ as $N\to\infty$ for any finite $y=n/N$.
In addition to that, close to the singularity at $y=0$ 
\begin{equation}
 \frac{s_{n,N-1}}{q_n} = \sum_{k=n}^{N-1} \frac{q_k}{q_n}
 = 1 + \frac{d}{b} + \left(\frac{d}{b}\right)^2 + \dots = \frac{b}{b-d} ~.
\end{equation}
Thus, inserting the above asymptotic forms into Eq.~(\ref{tfixfinal}), the terms around $y=0$ have a contribution
\begin{equation}
 \frac{d}{b-d} \sum_{n=1} \frac{1}{y} ~,
\end{equation}
where we do not specify the upper limit as the relevant contribution comes from the terms close to the lower limit
\begin{equation}
 \sum_{n=1} \frac{1}{y} \sim N \int\limits_{1/N} dy \frac{1}{y} \sim N\ln N ~.
\end{equation}
By taking into account the terms close to the other singularity at $y=1$ as well, where  
\begin{equation}
 \frac{s_{n,N-1}}{q_n} = \sum_{k=n}^{N-1} \frac{q_k}{q_n}
 = 1 + \frac{c}{a} + \left(\frac{c}{a}\right)^2 + \dots = \frac{a}{a-c} ~,
\end{equation}
we arrive the large $N$ asymptotic expression for the fixation time
\begin{equation}
 t_\mathrm{fix} \approx \left(\frac{a}{a-c}+\frac{d}{b-d}\right) N\ln N \sim N\ln N ~.
\end{equation}
According to numerical evaluation of Eq.~(\ref{tfixfinal}), the $N\ln N$ part is asymptotically exact but the amplitude is probably just a very good approximation.

Now, the $B\longleftarrow A$ case can be easily obtained by interchanging the players $A\leftrightarrows B$ in the previous case, and by noting that the fixation time is always the same for both players as stated in Eq.(\ref{equal}). Hence 
\begin{equation}
  t_\mathrm{fix} \approx \left(\frac{a}{c-a}+\frac{d}{d-b}\right) N\ln N  \sim N\ln N ~.
\end{equation}

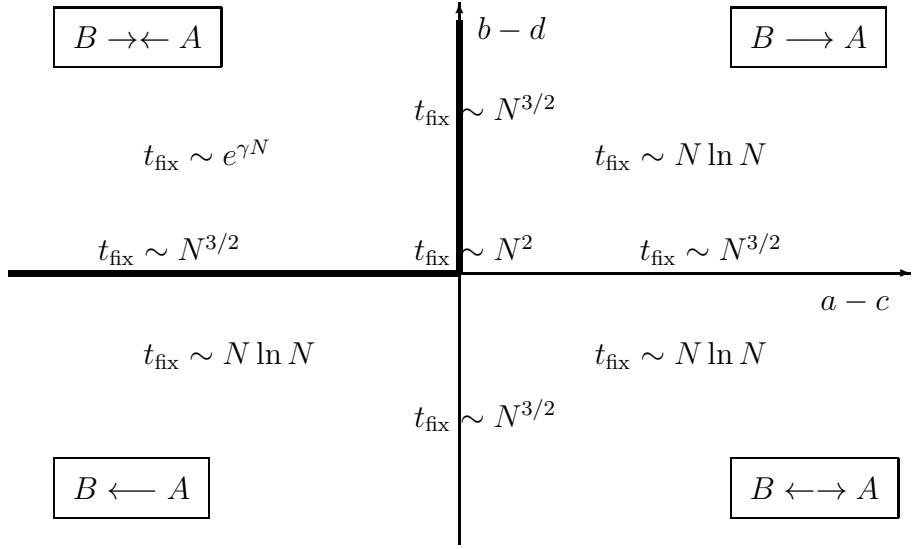
\begin{figure}
\setlength{\unitlength}{1.2mm}
\begin{center}
\begin{picture}(100,60)(0,0)
\put(0,30){\vector(1,0){100}}
\put(90,26){$a-c$}
\put(50,0){\vector(0,1){60}}
\linethickness{2pt}
\put(0,30){\line(1,0){50}}
\put(50,30){\line(0,1){28}}
\put(52,56){$b-d$}
\put(5,55){\framebox{$B\rightarrow\leftarrow A$}}
\put(5,5){\framebox{$B\longleftarrow A$}}
\put(80,55){\framebox{$B\longrightarrow A$}}
\put(80,5){\framebox{$B\leftarrow\rightarrow A$}}
\put(15,42){$t_\mathrm{fix}\sim e^{\gamma N}$}
\put(15,20){$t_\mathrm{fix}\sim N\ln N$}
\put(65,42){$t_\mathrm{fix}\sim N\ln N$}
\put(65,20){$t_\mathrm{fix}\sim N\ln N$}
\put(45,31.5){$t_\mathrm{fix}\sim N^2$}
\put(10,31.5){$t_\mathrm{fix}\sim N^{3/2}$}
\put(70,31.5){$t_\mathrm{fix}\sim N^{3/2}$}
\put(45,47){$t_\mathrm{fix}\sim N^{3/2}$}
\put(45,13){$t_\mathrm{fix}\sim N^{3/2}$}
\end{picture}
\end{center}
\caption{Graphical illustration of the different scenarios for the fixation time. The large $N$ behavior of the fixation time discussed in Table~\ref{sum} is displayed (the value of $\gamma>0$ can be read out from Table~\ref{sum}). The thick lines refer to relevant finite size corrections to the $N\to\infty$ behavior.}
\label{diagtime}
\end{figure}

Next we consider the $B\rightarrow\leftarrow A$ case for $\tilde q(1)<1$. 
The quantity which develops the relevant singularity in the $N\to\infty$ limit is $1/q_n$. It is indeed singular at $y^*$ of Eq.~(\ref{ystar}), since $q_n=\tilde q(y)^N$, with $y=n/N$, and as $q(y)$ takes its minimum $q^*=q(y^*)$ at $y^*$. Similarly to the previous case, $s_{0,n-1}/s_{0,N-1}\to1$ as $N\to\infty$,  for any finite $y=n/N$. The term $s_{n,N-1}$ gains the relevant contribution around $y=1$ as long as $\tilde q(n/N)<\tilde q(1)$, which is the case around the singularity as $q^*<\tilde q(1)$. This makes $s_{n,N-1}\approx q_{N-1}+q_{N-2}+\dots\sim\tilde q(1)^N$. As we showed in the previous case, $\lambda_n$ has only weak singularities at $y=0$ and $y=1$, leading to an $N\ln N$ contribution to $t_\mathrm{fix}$, which can be safely neglected in this case. Hence $\lambda_n$ can be considered a constant around $y^*$. Putting now all these asymptotic expressions back to Eq.~(\ref{tfixfinal}) we arrive at
\begin{equation}
 t_\mathrm{fix} \sim \sum_{n=1}^{N-1} \left(\frac{\tilde q(1)}{q^*}\right)^N \sim \sqrt{N}
 \left(\frac{\tilde q(1)}{q^*}\right)^N ~,
\end{equation}
where, at the last step, we used the method of steepest descent again. The case $B\rightarrow\leftarrow A$ case for $\tilde q(1)>1$ is entirely similar, with the only difference that in the $N\to\infty$ limit $s_{n,N-1}/s_{0,N-1}\to 1$, while $s_{0,n-1}\to b/(b-d)=$const around the singularity $y^*$ of $1/q_n$. Hence, in this case
\begin{equation}
 t_\mathrm{fix} \sim \sqrt{N} \left(\frac{1}{q^*}\right)^N ~.
\end{equation}

In the remaining $B\leftarrow\rightarrow A$ region we have three singularities in the expression of Eq.~(\ref{tfixfinal}) for $t_\mathrm{fix}$ at $y=0, y^*$ and 1. The contribution coming from the terms close to $y=0$ and $y=1$ is $\sim N\ln N$, analogously to the $B\longrightarrow A$ case. We provide a crude argument that the contribution from the singularity at $y^*$ is smaller, namely $\sim N$. 

Let us approximate quantities in Eq.~(\ref{tfixfinal}) around the singularity $y^*$ up to second order in $\delta=y-y^*$. Around its maximum $\tilde q(\delta)\approx q^*(1-\beta\delta^2)$, with $\beta>0$. This makes
\begin{equation}
 q_n(\delta) \approx q^{*N} e^{-N\beta\delta^2} ~,
 \label{little}
\end{equation}
hence $s_{0,N-1}\sim\sqrt N q^{*N}$, and in this approximation $s_{0,n^*-1}\approx s_{n^*,N-1}\approx s_{0,N-1}/2$, where $n^*=Ny^*$. Integrating Eq.(\ref{little}) for small $\delta\ll 1/\sqrt{N\beta}$, we arrive at
\begin{equation}
 s_{0,n-1} \sim \sqrt N q^{*N} \left( 1+ \beta' \sqrt N \delta + O(\delta^3) \right) ~, 
\end{equation}
where $\beta'>0$ is a constant, and the same for $s_{n,N-1}$, with $-\beta'$ instead of $\beta'$. Putting these asymptotic expressions into Eq.~(\ref{tfixfinal}) and noting that $\lambda_n$ can be considered as constant around $y^*$, we get
\begin{equation}
 t_\mathrm{fix} \sim \sqrt N \sum_n 1+ \beta'' N \delta^2 
 \sim N^{3/2} \int d\delta~ e^{-\beta''N\delta^2} \sim N ~,
\end{equation}
where $\beta''>0$ is another constant.
Together with the contributions from $y=0$ and 1 the asymptotic behavior is
\begin{equation}
 t_\mathrm{fix} \sim N\ln N ~.
\end{equation}

\begin{table}[htdp]
\begin{center}
\begin{tabular}{l||c|c|c}
 &  $\epsilon_1^+$ & $\epsilon_{N-1} $ & $t_\mathrm{fix}$\\
\hline\hline
$B\longrightarrow A$ ($a>c$, $b>d$)&  $1-d/b$ & $\sim \tilde q(1)^N$ & $\sim N\ln N$\\
\hline
$B\longleftarrow A$ ($a<c$, $b<d$) & $\sim \tilde q(1)^{-N}$ & $1-a/c$ & $\sim N\ln N$\\
\hline
$B\leftarrow\rightarrow A$ ($a>c$, $b<d$) & $\sim (1/q^*)^N$ & $\sim (\tilde q(1)/q^*)^N$ & $\sim N\ln N$\\
\hline
$B\rightarrow\leftarrow A$ ($a<c$, $b>d$), 
     ~~$\tilde q(1)<1$ & $1-d/b$ & $\sim \tilde q(1)^N$ & $\sim (\tilde q(1)/q^*)^N$\\
\cline{2-4}
\hfill $\tilde q(1)>1$ & $\sim \tilde q(1)^{-N}$ & $1-a/c$ & $\sim (1/q^*)^N$\\
\hline
$a=c, ~ b=d$ & $\sim 1/N$ & $\sim 1/N$ & $\sim N^2$\\
\hline
($a=c$, $b>d$) or  ($b=d$, $a<c$) & $\sim \tilde q(1)^{-N}$ & $\sim 1/\sqrt{N}$ & $\sim N^{3/2}$\\
\hline
($a=c$, $b<d$) or  ($b=d$, $a>c$) & $\sim 1/\sqrt{N}$ & $\sim \tilde q(1)^{N}$ & $\sim N^{3/2}$\\
\hline
\end{tabular}
\end{center}
\caption{Fixation probabilities and times in the $N\to\infty$ limit for different scenarios. Note, that due to the specific values of $\tilde q(1)$ and $q^*$, when the above quantities depend exponentially on $N$, the fixation probabilities are exponentially {\it small}, while the fixation times are exponentially {\it large}.}
\label{sum}
\end{table}

We complete this section as well with investigating the asymptotic behavior on the axes. In the origin ($a=c$ and $b=d$) we asymptotically recover the results for the neutral game (Sec.~\ref{neutral}).
As $g_i/f_i\approx q_i\approx 1$, $s_{n,m}\approx m-n$ and $\lambda_i\approx y(1-y)$ we obtain $t_\mathrm{fix}\sim N^2$.

For $a=c$ and $b>d$ the maximum of $\tilde q(y)$ is at 0 with $\tilde q'(0)<0$, hence similarly to the $B\longrightarrow A$ case, the terms around $y=0$ have an $N\ln N$ contribution. The relevant contribution though comes from the terms around $y=1$ where $\tilde q'(1)=0$. Around $y=1$ the ratio $s_{0,n-1}/s_{0,N-1}\approx 1$ and $\lambda_n\approx \delta$ with $\delta=1-y=1-n/N$. By approximating $\tilde q_n\approx \tilde q(1)(1+\beta\delta^2)$ up to second order with $\beta>0$, we obtain 
\begin{equation}
s_{n,N-1}\approx N \tilde q(1)^N \delta (1+\frac{\beta n}{3}\delta^2) ~.
\end{equation}
Inserting these terms into Eq.~(\ref{tfixfinal}) and using the steepest descent method we arrive at the algebraically decaying asymptotic behavior $t_\mathrm{fix}\sim N^{3/2}$. The same behavior occurs for $b=d$ and $a<c$, as it follows by interchanging the players.

The $b=d$ and $a>c$ case goes along similar lines. The relevant contribution in this case comes from the terms around $y=0$ where $\tilde q(y)$ takes its maximum and $\tilde q'(0)=0$. In the leading order $\lambda_n\approx y$, $\tilde q_n\approx 1-\beta y^2$ with $\beta>0$, $s_{n,N-1}/s_{0,N-1}\approx 1$, and
\begin{equation}
s_{0,n-1}\approx N y (1-\frac{\beta n}{3}y^2) ~.
\end{equation}
Using Eq.~(\ref{tfixfinal}) we arrive at the same algebraic asymptotic behavior $t_\mathrm{fix}\sim N^{3/2}$. Interchanging the strategies leads to the same result also for $a=c$ and $b<d$.

These results for the asymptotic behavior of the absorption times are summarized in Table~\ref{sum} and also depicted in Fig.~\ref{diagtime}.

\section{Numerical examples}
\label{numer}

\begin{figure}
\centering
\includegraphics[scale=0.8]{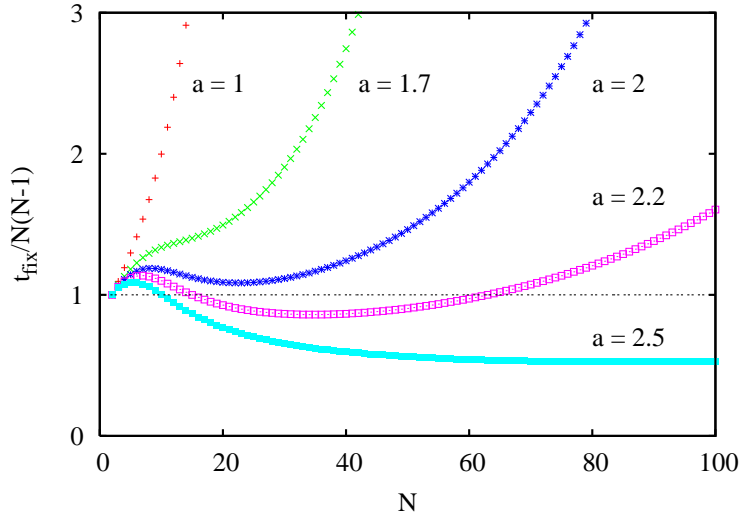}
\caption{Illustrative examples for the relative fixation time as a function of population size when there is an ESS with coexisting strategies in the infinite population size limit, $B\rightarrow\leftarrow A$. 
The payoff matrix is \mymatrix{$a$}{4}{3}{1}, where parameter $a$ varies as indicated in the figure.}
\label{coex}
\end{figure}

In this section we investigate the population size $N$ dependence of the fixation time $t_\mathrm{fix}$ based on numerical evaluation of the exact formula  Eq.~(\ref{tfixfinal}). 
For large $N$ values the asymptotic forms summarized in Table~\ref{sum} are recovered. 
Relevant deviations from these asymptotic forms are present close to the boundaries of the two types (exponential and $N\ln N$) of behavior, namely at the boundaries of the $B\rightarrow\leftarrow A$ region, marked with thick lines in Fig.~(\ref{diagtime}).

As a typical example for large finite size effects we consider payoff matrices
of the form $\begin{pmatrix} a~&4\\ 3&1\end{pmatrix}$ with $a<3$, which then belong to the $B\rightarrow\leftarrow A$ case. For $a=1$, one sees the exponentially growing fixation time in Fig.~\ref{coex}. As $a\to3$, that is as we approach the boundary $a-c=0$ in Fig.~\ref{diagtime}, the exponential growth can be observed only for larger population sizes. As it is shown  in Fig.~\ref{coex} for small population sizes, $t_\mathrm{fix}$ can even be smaller than $N(N-1)$, the fixation time for neutral games. For smaller $c-a$ values the exponential growth of the fixation time starts at larger $N$ values.
The same finite size behavior is observed as we approach the $b-d=0$ boundary from the same $B\rightarrow\leftarrow A$ region.

\begin{figure}
\centering
\includegraphics[scale=0.8]{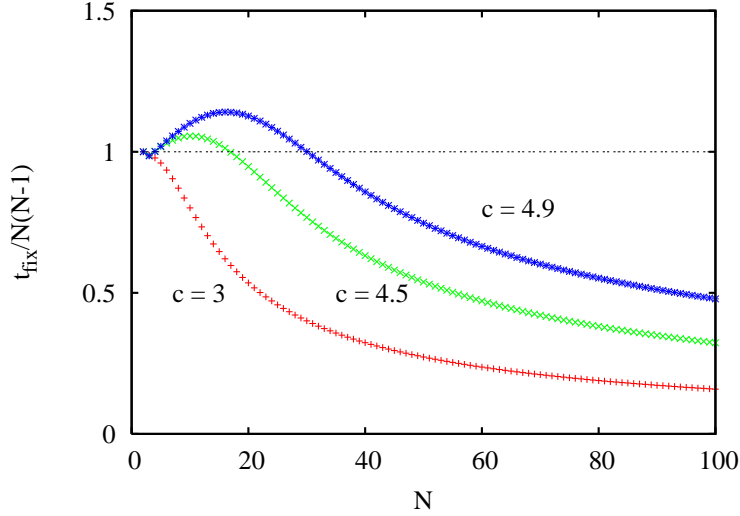}
\caption{Illustrative examples for the relative fixation time as a function of population size when there are only one ESS-s in the infinite population size limit, $B\longrightarrow A$. 
The payoff matrix is \mymatrix{5}{2}{$c$}{1}, where parameter $c$ varies as indicated in the figure.}
\label{domin}
\end{figure}

Finite size effects are also relevant in the $B\longrightarrow A$ region when approaching the $a-c=0$ boundary. In this case the asymptotically slow ($\sim N\ln N$) fixation time can exceed the neutral value $N(N-1)$ for small population sizes $N$. This behavior can be observed on an example in Fig.~\ref{domin}. Again, in general, for smaller $a-c$ values the fixation time exceeds that of the neutral game up to larger values of $N$. The same scenario occurs in the $B\longleftarrow A$ region as $d-b\to0$.
Conversely, finite size effects are not relevant at the boundaries of the $B\leftarrow\rightarrow A$ region.
This fact is not surprising as the large $N$ behavior of the fixation time is also $\sim N\ln N$ in the neighboring domains, and even on the boundaries it is still smaller ($N^{3/2}$) than in the neutral game.

Now we make three conjectures about the fixation time for finite $N$ values. These conjectures are based on numerical evaluation of Eq.~(\ref{tfixfinal}) for numerous ($\sim 50000$) random (all elements are from the interval $[0,1]$) payoff matrices in for each system sizes from $N = 2$ to $N= 100$.
{\bf First conjecture:} ${t}_\mathrm{fix}(N)\le N(N-1)$ for {\it all} values of $N$ in the $B\leftarrow\rightarrow A$ region. The equality (${t}_\mathrm{fix}(N)=N(N-1)$) was only observed in the trivial $N=2$ case. {\bf Second conjecture:} if $\epsilon_{N-1}>1/N$ and $\epsilon^+_1>1/N$ then ${t}_\mathrm{fix}(N) \ge N(N-1)$. {\bf Third conjecture:} if $f_1-g_1 < 0$ and $f_{N-1}-g_{N-1} > 0$ then ${t}_\mathrm{fix}(N) \le N(N-1)$.  

Based on the above conjectures we can predict the speed of the process in three out of the five evolutionary scenarios established in \citep{nowak04}.
According to our 2. conjecture if mutual invasion and mutual fixation is favored then fixation was experienced to be slow without exception (case 2. in Table~\ref{combi}).
Thus a state with coexisting strategies is likely to be observed in this situation. This is a qualitatively new situation not present in \citep{nowak04}. If selection opposes mutual invasion then fixation is fast (3. conjecture),  independently whether selection favors fixation or not (cases 4. 5. in Table~\ref{combi}.). This conjecture is also supported by the large $N$ asymptotics (see Fig. \ref{diagtime}.).

\section{Discussion}
\label{disco}

We have used a recently introduced frequency dependent Moran process to study game dynamics in finite populations. The probability that a single mutant strategy $A$ can fixate in a population consisting $N-1$ individuals with strategy $B$ has been known \citep{karlin, Nowak, nowak04}. The stationary distribution for this system is studied recently. It is shown numerically that this distribution can deviate significantly from the Gaussian one \citep{Claussen}. We studied the average time to fixation in this paper. We have shown that the fixation time is identical for the two strategies for a given game independently of the elements of the payoff matrix and the population size, that is $t_\mathrm{fix}(N)=t_{N-1}=t^+_1$. In fact this property is necessary for fixation probabilities to meaningfully characterize selection scenarios. As an example consider a process where the fixation probabilities are the same for the two strategies ($\epsilon_{N-1} = \epsilon_1^+$). Then in order to conclude that the population is observed with the same probability in both absorbing states \citep{Fudenberg} it is necessary to know that both absorption times are the same.

Since the exact results for the fixation time and fixation probabilities are overly complicated as a function of the payoff matrix, we also determined their large $N$ asymptotic behavior. 
We have shown that $t_\mathrm{fix} \sim N \ln N$ if there is at least one pure ESS in the infinite population size limit. Thus, the fixation time becomes negligibly smaller than that of the neutral game as the population size goes to infinity ($\ln N/(N-1) \rightarrow 0$ if $N\rightarrow \infty$). Interestingly, fixation is fast even if the system is bistable (there are two ESS-s), although to arrive from the neighborhood of one of the stable states to the other one is impossible in the standard deterministic replicator dynamics of the infinite populations \citep{Hofbauer}. On the other hand we have shown that the fixation probability of a mutant is exponentially small in the large population size limit.
A particularly surprising result is that, as the population size grows, the fixation probability of a dominating strategy tends to a constant less than one. This probability depends only on the relative payoff of the non-dominant strategy against itself to the payoff of the dominant strategy against the non dominant one (see Table \ref{sum}). Consequently, in real stochastic populations (even in very large ones) the probability of fixation of a dominant mutant can be very small when the mutant strategy receives only slightly bigger payoff against the resident strategy than the resident does against itself. One can gain an intuitive insight into this result.  When the mutant is rare, it meets with the resident strategy much more frequently than with another rare mutants. If the mutant has only a small advantage in this state then it can extinct easily. This is why the fixation probability of the dominant mutant is small in this case. In the reverse case, if the mutant plays much more effectively against the resident strategy than the resident itself, then the rare mutant will spread with high probability, so their fixation probability is close to one. 

If strategies $A$ and $B$ coexist in the infinite size deterministic model then the fixation time increases exponentially with $N$, that is the fixation time tends to infinity compared to the neutral game (($e^N/[\sqrt{N}(N-1)]\rightarrow \infty$).  
Hence in the large population limit $A$ and $B$ remain practically in a state of coexisting strategies. Depending on a complex relation among the payoff elements ($\tilde q(1)\lessgtr 1$) one of the strategies fixates with a constant probability while the fixation probability of the other decreases exponentially with $N$ (Table \ref{sum}.). While fixation is slow in this case, it is probable for a rare mutant that plays effectively against the resident strategy (see above).      

Our results about the bistable case have some consequences in the context of the origin of cooperation.
The origin of cooperative behavior among selfish individuals remains an intriguing problem of evolutionary biology \citep{axel,Szathmary, Szabo, Nowak-Sigmund}.
Using the iterated prisoner's dilemma game as a conceptual model \citep{axel} Nowak et al. \citep{Nowak} considered a game where a cooperative (Tit-for-Tat) strategy plays against an always defective strategy (AllD). In this situation both the Tit-for-Tat and AllD strategies are stable fixed points of the classical replicator dynamics. Consequently, although a purely cooperative population is a stable state,  Tit-for-Tat can not spread in a defective population. However, this argument is true only for infinite deterministic populations. Nowak et al. demonstrated, using the model studied in the present paper, that a single mutant cooperative Tit-for-Tat can replace the $N-1$ AllD players with high probability, for appropriate $N$ values (not too small, not too large). This means in our context, that the fixation probability of Tit-for-Tat could be four or five times greater than it would be in the neutral case ($1/N$). On the other hand, replacement of Tit-for-Tat by AllD is unlikely since  fixation probability of AllD is smaller than $1/N$ for appropriate $N$-s. That is, selection follow the scenario (4) in Table \ref{combi}, where Tit-for-Tat corresponds to strategy $B$. Nevertheless, the speed of the fixation of Tit-for-Tat remained an open question.
According to our third conjecture and our analysis of the large $N$ limit, however, if selection opposes mutual invasion then the fixation is fast (cases 4 and 5 in Table~\ref{combi}). Consequently, if the population size is appropriate for the fixation of the cooperative strategy, then this fixation will be fast, making the spread and fixation of the cooperative strategy more likely in finite populations.

\section*{Acknowledgments}
TA is thankful to S. Redner, R. K. P. Zia, and especially to P. L. Krapivsky for useful discussions.
TA also gratefully acknowledges financial support from the Swiss
National Science Foundation under the fellowship 8220-067591.
This work was founded in part by OTKA Grants T037726, T043734 and T049692.

\section*{Appendix}

In Eq.(\ref{equal}) we arrived at the surprising result that the fixation times are always the same for both species. In this section we provide a simple independent argument to support this result.
Say we have a particular walk $s$, which starts from site 1 at time 0, and arrives at site $N$ for the first time at time $T(s)$, without having stepped on site 0. It is obvious that this walk has to step through each bond to the right (towards site $N$) at least once. It is also clear that the walk steps one more times to the right then to the left through each bond, as eventually the walk ends up at site $N$. This means that the probability $P(s)$ of the walk $s$ can be written as
\begin{equation}
 P(s) = \lambda_{N-1} \prod_{i=1}^{N-2} 
 \lambda_i^{1+\alpha_i} \mu_{i+1}^{\alpha_i} \prod_{i=1}^{N-1} \nu_i^{\beta_i} = 
 \pi(s) \prod_{i=1}^{N-1} \lambda_i ~,
 \end{equation}
with $\alpha_i$ and $\beta_i$ being non-negative integers, $\nu_i=1-\mu_i-\lambda_i$, and we used the notation
\begin{equation}
 \pi(s) = \prod_{i=1}^{N-2} (\lambda_i \mu_{i+1})^{\alpha_i} \prod_{i=1}^{N-1} \nu_i^{\beta_i} ~.
\end{equation}
Now for each walk $s$ we can uniquely construct a walk $\tilde s$, which walks backward from site $N-1$ to site 0 on the path of $s$, that is $\tilde s_t = N-s_{T(s)-1-t}$ for $0\le t<T(s)$, and $\tilde s_{T(s)} = N$. The duration of the two walks are obviously the same $T(s) = T(\tilde s)$. 
Since $\tilde s$ steps through each bond backward compared to $s$, its probability can be obtained by interchanging all $\lambda_i$ by $\mu_{i+1}$ and vice versa, apart from the last steps of each walk, where we have to change $\lambda_N$ to $\mu_1$, that is
\begin{equation}
 P(\tilde s) = \mu_1 \prod_{i=1}^{N-2} 
 \lambda_{i}^{\alpha_i} \mu_{i+1}^{1+\alpha_i} \prod_{i=1}^{N-1} \nu_i^{\beta_i} = 
 \pi(s) \prod_{i=1}^{N-1} \mu_i
\end{equation}
In order to calculate the average conditional first passage time $t_1^+$ we have to sum over all path $s$, which starts from site $1$ and exits at site $N$, that is
\begin{equation}
 t_1^+ = \frac{\sum\limits_{\{ s\}} T(s) P(s)}{\sum\limits_{\{ s\}} P(s)}
 = \frac{\sum\limits_{\{ s\}} T(s) \pi(s)}{\sum\limits_{\{ s\}} \pi(s)} ~.
\end{equation}
Similarly, by summing over the walks which start at site $N-1$ end exit at site $0$ we arrive at
\begin{equation}
 t_{N-1} = \frac{\sum\limits_{\{ \tilde s\}} T(\tilde s) P( \tilde s)}{\sum\limits_{\{ \tilde s\}} P( \tilde s)}
 = \frac{\sum\limits_{\{ \tilde s\}} T(s) \pi(s)}{\sum\limits_{\{ \tilde s\}} \pi(s)} ~.
\end{equation}
Since there is a one to one correspondence between $\{ s\}$ and $\{ \tilde s\}$, the two sums are the same, that is $t_1^+ = t_{N-1}$, what we wanted to demonstrate.

It is interesting to note that the same argument also applies to the conditional first passage probabilities
\begin{equation}
 P(t_1^+=\tau) = \frac{\sum\limits_{\{ s\}} \delta_{T(s),\tau} P(s)}{\sum\limits_{\{ s\}} P(s)}
= \frac{\sum\limits_{\{ \tilde s\}} \delta_{T(\tilde s), \tau} P( \tilde s)}{\sum\limits_{\{ \tilde s\}} P( \tilde s)}
 = P(t_{N-1}=\tau) 
\end{equation}
that is not only the average first passage time, but the whole conditional first passage probability to arrive at site $N$ from site 1 in time $\tau$ is the same as that to arrive at site 0 from site $N-1$ in time $\tau$.


\begin{thebibliography}{99}

\bibitem[Axelrod, 1981]{axel} Axelrod, R. \& Hamilton, W. D. (1981). The evolution of cooperation. \emph{Science} \textbf{211,} 1390-1396.

\bibitem[Claussen \& Traulsen, 2005]{Claussen} Claussen, J. C. \& Traulsen, A. (2005).
Non-Gaussian fluctuations arising from finite populations: Exact results for the evolutionary Moran process. \emph{Phys. Rev. E.} \textbf{71,} 025101(R). 

\bibitem[Fisher, 1988]{fisher} Fisher, M. E. (1988). Diffusion from an entrance to an exit. IBM J. Res.\ Develop.\ \textbf{32,} 76-81.

\bibitem[Fudenberg et al., 2005]{Fudenberg} Fudenberg, D., Imhof, L., Nowak, M. A. \& Taylor, C. (2005). Stochastic evolution as generalized Moran process (preprint).

\bibitem[Hofbauer \& Sigmund, 2003]{Hofbauer} Hofbauer, J. \& Sigmund, K. (2003). Evolutionary game dynamics. \emph{Bull. Am. Math. Soc.} \textbf{40,} 479--519.

\bibitem[Karlin \& {Taylor, H.}, 1975]{karlin} Karlin, S. and Taylor, H. (1975). {\it A First Course in Stochastic Processes}, Academic Press. 

\bibitem[Landauer \& B\"uttiker, 1987]{landauer} Landauer, R. and B\"uttiker, M. (1987). Diffusive traversal time: Effective area in magnetically induced interference. \emph{Phys.\ Rev. B} \textbf{36,} 6255-6260.

\bibitem[Lieberman et al., 2005]{Lieberman} Lieberman E., Hauert C. \& Nowak M. A. (2005). Evolutionary dynamics on graphs. \emph{Nature} \textbf{433,} 312--316.

\bibitem[Maynard Smith, 1988]{Maynard2} Maynard Smith, J. (1988). Can mixed strategy be stable in a finite population?  \emph{J. Theor. Biol.} \textbf{130,} 247--251.

\bibitem[Maxnard Smith, 1974]{Maynard} Maynard Smith, J. (1974). The theory of games and the evolution of animal conflicts. \emph{J. of Theor. Biol.} \textbf{47,} 209--221.

\bibitem[Maynard Smith \& Szathm\'ary, 1995]{Szathmary} Maynard Smith, J. \& Szathm\'ary, E. (1995). {\it The Major Transitions in Evolution}, Oxford: Freeman, Spektrum.

\bibitem[Moran, 1962]{Moran} Moran, P. A. P. (1962). {\it The Statistical Processes of Evolutionary Theory}, Oxford: Clarendon Press.

\bibitem[Neil, 2004]{Neill} Neill, D. B. (2004). Evolutionary stability for large populations. \emph{J. of Theor. Biol.} \textbf{227,} 397--401.

\bibitem[Nowak \& Sigmund, 2004]{Nowak-Sigmund} Nowak, M. A.,\& Sigmund, K. (2004). Evolutionary dynamics of biologically games. \emph{Science} \textbf{303,} 793--799.

\bibitem[Nowak et al., 2004]{Nowak} Nowak, M. A., Sasaki, A., Taylor, C. \& Fudenberg, D. (2004). Emergence of cooperation and evolutionary stability in finite populations. \emph{Nature} \textbf{428,} 646--650.

\bibitem[Redner, 2001]{fpp} Redner, S., (2001). {\it A Guide to First-Passage Processes}, New York: Cambridge University Press.

\bibitem[{Taylor, C.} et al., 2004]{nowak04} Taylor, C., Fudenberg, D. Sasaki, A. \& Nowak, M. A. (2004). Evolutionary game dynamics in finite populations. \emph{Bull. Math. Biol.} \textbf{66,} 1621--1644.

\bibitem[Schaffer, 1988]{Schaffer} Schaffer, M. (1988). Evolutionary stable strategies for a finite population and a variable contest size. \emph{J. Theor. Biol.} \textbf{132,} 469--478.

\bibitem[Szab\'o et al., 2004]{Szabo} Szab\'o, G., Vukov, J., Szolnoki, A. (2005). Phase diagrams for an evolutionary PrisonerÕs Dilemma game on two-dimensional lattices, e-print cond-mat/0506433.

\bibitem[van Kampen, 1997]{vankampen} van Kampen, N. G. (1997). {\it Stochastic Processes in Physics and Chemistry}, 2nd ed. Amsterdam: North-Holland.

\bibitem[Wild \& {Taylor, P. D.}, 2004]{wild} Wild, G. and Taylor, P. D. (2004). Fitness and evolutionary stability in game theoretic models of finite populations. \emph{Proc.\ R. Soc.\ Lond.\ B} \textbf{271,} 2345-2349.

\end{thebibliography}
\end{document}